\documentclass[aps,reprint,prd]{revtex4-2}
\usepackage[T1,T2A]{fontenc}
\usepackage{hyperref,amsmath,amssymb,mathtools,mleftright}
\usepackage[german,french,czech,latin,british]{babel}
\usepackage{CJKutf8}
\usepackage[utf8]{inputenc}
\usepackage[final,spacing]{microtype}
\microtypecontext{spacing=nonfrench}

\hypersetup{
    colorlinks=true,
    citecolor=red,
    linkcolor=blue,
    filecolor=cyan,      
    urlcolor=magenta,
}

\hypersetup {
    pdftitle = {Discrete 𝑝‐form symmetry and higher Coulomb phases},
    pdfauthor = {Leron Borsten and Hyungrok Kim},
    pdfsubject = {hep-th, math-ph, hep-lat},
    pdflang = {en-GB},
    pdfkeywords = {higher form symmetry, Coulomb phase, adjusted higher gauge theory, Villain model, BF model},
}

\begin{document}
\title{Discrete \(p\)-Form Symmetry and Higher Coulomb Phases in Various Theories}
\author{Leron Borsten}
\affiliation{Blackett Laboratory, Imperial College London, London SW7 2AZ, United Kingdom}
\affiliation{Department of Physics, Astronomy and Mathematics, University of Hertfordshire, Hatfield, Hertfordshire AL10 9AB, United Kingdom}
\email{l.borsten@herts.ac.uk}
\author{Hyungrok Kim}
\affiliation{Department of Physics, Astronomy and Mathematics, University of Hertfordshire, Hatfield, Hertfordshire AL10 9AB, United Kingdom{}{}}
\email{h.kim2@herts.ac.uk}
\begin{abstract}
We argue that a field theory with a \(\mathbb Z_N\) \(p\)-form symmetry generically admits, in addition to a Higgs phase  and a `confining' phase, a Coulomb phase in which the infrared theory contains Abelian \(p\)-form electrodynamics, similar to the behavior of Yang--Mills theory coupled to adjoint or fundamental matter.
We illustrate our claim with continuum and lattice examples including  2-group higher gauge theory, effective $p$-form $BF$ theories and  lattice $p$-form electrodynamics. 
\end{abstract}
\maketitle
\tableofcontents

\section{Introduction}

\subsection{Summary} Global $p$-form symmetries generalize  familiar (i.e.~$0$-form) global symmetries and entail similarly important physical consequences; a prominent example is given by the role  one-form  symmetries  have played in our understanding of confinement.  Here, we show that for any theory with a \(\mathbb Z_N\) \(p\)-form symmetry there are monopole \((d-3-p)\)-branes that carry magnetic \(p\)-form charge under the  symmetry, which source \(N\) units of \((d-2-p)\)-brane centre vortices. Consequently, all such theories are argued to have Coulomb, Higgs and confining  phases for \(d\ge p+3\). We give lattice and continuum higher gauge theory examples exhibiting  this behavior. This generalizes the conclusions of \cite{Nguyen:2024ikq} for \(\mathbb Z_N\) one-form symmetries and includes an, to the best of our knowledge, original understanding how adjoint matter couples to a higher gauge theory. In the process, we explain how higher gauge theory entails  $\infty$-group generalized global symmetries, giving an explicit 2-group example, in direct analogy to the one-form centre symmetry of Yang--Mills theory.  This complements previous constructions of $\infty$-group global symmetries, which typically proceed in the other direction starting from the higher global symmetries themselves, as realized by extended topological operators, and then coupling to higher form (background) gauge potentials  \cite{Gaiotto:2014kfa}.

 \subsection{Background}   Although  global $p$-form symmetries (also referred to as `generalized global symmetries' or `higher form symmetries') have previously appeared in various guises  in various contexts, for example in  \cite{Polyakov:1976fu,Polyakov:1975rs,tHooft:1977nqb,Kovner:1992pu,deWildPropitius:1995hk,Alford:1990fc, Bucher:1991bc, Alford:1992yx, Pantev:2005rh,Pantev:2005zs,Hellerman:2006zs,Nussinov:2009zz,Kapustin:2013uxa}, they were only  systematically characterized in the   seminal work of \cite{Gaiotto:2014kfa}. Subsequently,  such higher form symmetries have become an increasingly central aspect of  gauge theory, as reviewed in \cite{Cordova:2022ruw,Brennan:2023mmt,Gomes:2023ahz,Bhardwaj:2023kri,Luo:2023ive}, and found to be closely related to topological phases,  e.g.\ \cite{Gukov:2013zka,Kapustin:2013uxa,Barkeshli:2014cna,Gukov:2014gja,Nguyen:2024ikq}.

A well-known  example of a one-form symmetry by another name is the $\mathbb Z_N$ `centre symmetry' of Polyakov and 't~Hooft \cite{Polyakov:1976fu,Polyakov:1975rs,tHooft:1977nqb}\footnote{Of course, this predates the present understanding of generalized global symmetries; Polyakov and 't~Hooft  did not articulate the Yang--Mills centre symmetry in terms of  a one-form symmetry. }.  Specifically, an \(\operatorname{SU}(N)\) gauge theory with adjoint matter admits a \(\mathbb Z_N\) one-form symmetry and has a Higgs phase, a Coulomb phase, and a confining phase; in particular,  the \(\mathbb Z_N\) one-form symmetry plays a crucial role in confinement. See  reviews  \cite{Ogilvie:2012is,Holland:2000uj} and the references therein. Given this insight, it  is natural to ask if the relationship between the one-form symmetry and the phases of Yang--Mills theory generalizes.

 Indeed,   it was recently shown \cite{Nguyen:2024ikq} that such phases occur generically in \emph{any} \(d\ge4\)-dimensional quantum field theory with a \(\mathbb Z_N\) one-form symmetry. In contrast to the dual-superconductor picture mediated by magnetic monopole condensation, in this scenario confinement is driven by \emph{centre vortices} (which are \((d-3)\)-branes) sourced by  magnetic monopoles (which are \((d-4)\)-branes) carrying \(N\) units of charge. Depending on which of these two kinds of defects proliferate, the theory exhibits Higgs, Coulomb, or confining behavior (cf.~also \cite{Hayashi:2024yjc,Hayashi:2024psa}). In particular, in the Coulomb phase, one generically obtains a propagating \(\operatorname U(1)\) Maxwell field in the low-energy effective theory. This prediction has been confirmed numerically in lattice models \cite{Giansiracusa:2025hfj}.
Assuming a \(\mathbb Z_N\) symmetry (which remains  unbroken in the Coulomb phase), one can microscopically picture the situation as follows: there exists  particles (vortices) carrying unit  \(\mathbb Z\)-valued charge in a sea of light particles (monopoles) that instead carry \(N\) units of charge. Because of the presence of this sea, the \(\mathbb Z\)-valued conserved charge is broken to a \(\mathbb Z_N\)-valued conserved charge. In this case, the \(\mathbb Z\)-valued vortices then couple (magnetically, say) to a Maxwell field that provides the conservation law for the charge.

\subsection{Results}  In this paper, we discuss the higher-dimensional analogue of the above scenario, where the one-form symmetry is generalized to a $p$-form symmetry. That is, in any theory with a \(\mathbb Z_N\) \(p\)-form symmetry, we have  monopole \((d-3-p)\)-branes  that carry magnetic charge under the \(p\)-form symmetry, and these source \(N\) units of centre vortices, the  \((d-2-p)\)-branes. The behavior of the system then depends on which of these two defects proliferate (i.e.\ become light). 
Since each monopole is attached to \(N\) centre vortices, it is impossible for a monopole to be light while centre vortices are heavy. Thus, we have three possibilities:
\begin{itemize}
\item When both monopoles and centre vortices are light and hence proliferate, the system is in a confining phase: the effective potential between two \((d-3)\)-dimensional endpoints of the monopoles increases linearly with respect to the distance, analogously to the area law of Wilson loops, and thus provide a higher-dimensional analogue of QCD strings between quark--antiquark pairs.
\item When both monopoles and centre vortices are heavy, the system is in a Higgs phase. The system is then described by an effective \(BF\) model in the infrared whose symmetry matches the \(\mathbb Z_N\) \(p\)-form symmetry of the ultraviolet theory.
\item When the centre vortices are light but the monopoles are heavy, the system develops a light \(\operatorname U(1)\) \((d-2-p)\)-form electromagnetic field; this is a higher analogue of the Coulomb phase of gauge theories.
\end{itemize}
The mechanism in which the Coulomb phase develops an Abelian field is an instance of electromagnetic duality: the \(\mathbb Z_N\)-valued \(p\)-form gauge field dualizes to a \(\operatorname U(1)\) \((d-2-p)\)-form gauge field.

This paper is organized as follows. In \autoref{sec:adjusted}, we consider the phases of an adjusted higher gauge theory coupled to matter fields of various representations.
In \autoref{sec:bf}, we start from a \(BF\) model effective description of the Higgs phase and argue that natural deformations away from the Higgs phase yield Coulomb and unbroken phases.
Finally, in \autoref{sec:lattice}, we consider the emergence of the higher Coulomb phase in a Villain model of \(p\)-form electrodynamics on the lattice.

\section{Phases of adjusted higher gauge theory}\label{sec:adjusted}
Yang--Mills theory coupled to matter provides a paradigmatic example of the relation between one-form symmetry (the centre symmetry)  and phases. In particular, when adjoint matter coupled to the gauge boson acquires a vacuum expectation value, the theory exhibits a Coulomb phase, where the low-energy spectrum contains Maxwell theory (corresponding to the Cartan subgroup that is the stabilizer subgroup of the adjoint vacuum expectation value). On the other hand, when  matter sitting in a generic representation acquires a vacuum expectation value, the theory often exhibits a Higgs phase where all gauge symmetry has broken down. Finally, in the unbroken phase the strongly coupled theory often shows confinement.

Here we consider the analogous mechanism for higher gauge theory.  Higher gauge theory is a higher-form generalization (categorification) of Yang--Mills theory, where the gauge symmetry is described by a Lie \(\infty\)-group. See  \cite{Baez:2010ya, Borsten:2024gox} for reviews. Essentially, a higher gauge theory consists of one-form, two-form, three-form,\ldots local gauge potentials, $A,B, C\dotsc$ valued in a gauge homotopy Lie algebra (also known as an $L_\infty$-algebra). The corresponding  two-form, three-form, four-form,\ldots field strengths  and gauge transformations are determined by the categorified symmetry data.

For definiteness (and tractability),  we take the   Lie \(\infty\)-group to be given by a  strict Lie 2-group \(\mathcal G\). In this case, the higher gauge theory consists only of  a one-form gauge potential \(A\)  and a two-form gauge potential \(B\). Note that the strict 2-group \(\mathcal G\) is the  (generically non-Abelian) higher gauge (as opposed to global) symmetry of the theory. The theory is mostly fixed by the choice of $\mathcal G$, up to  an important subtly that implies for a \emph{dynamical and non-Abelian} higher gauge theory the data of the  gauge \(\infty\)-group alone is  insufficient.   This follows from the observation that for  a generic  non-Abelian gauge \(\infty\)-group one must impose the so-called fake flatness condition \cite{Breen:2001ie,Aschieri:2003mw,Baez:2004in,Baez:2005qu} in which one of the field strengths is constrained to zero for consistency of the gauge algebra. For topological or Abelian higher field theories, this is of no consequence. However, the fake-flatness condition  consigns aspiring   non-Abelian  theories to be topological or essentially Abelian \cite{Gastel:2018joi,Samann:2019eei}. This has led to erroneous assumption that there are no dynamical and genuinely non-Abelian higher gauge theories. However, a subset of Lie \(\infty\)-groups admit a structure of \emph{adjustment} \cite{Sati:2008eg,Sati:2009ic,Samann:2019eei,Schmidt:2019pks,Kim:2019owc,Borsten:2021ljb,Tellez-Dominguez:2023wwr,Fischer:2024vak,Gagliardo:2025oio,future:2024aa} (see \cite{Borsten:2024gox} for a review), which lifts the fake-flatness conditions and so allows for non-Abelian dynamical higher gauge theories. Thus, in order for our model to exhibit dynamics, we consider an adjusted higher gauge theory. 

\subsection{Review of adjusted higher gauge theory}
Let us consider an adjusted higher gauge theory associated to a strict 2-group \(\mathcal G\) modelled by a crossed module \((H\to G)\), for $H$ and  $G$  Lie groups. This class of (adjusted) higher gauge theories is treated in detail in \cite{Rist:2022hci,Borsten:2024gox}, with explicit examples and applications. The corresponding local higher gauge potentials consist of 
a \(\mathfrak h\)-valued two-form gauge potential \(B\)  and a  \(\mathfrak g\)-valued one-form potential \(A\),
\begin{equation}
B \in \Omega^2(M, \mathfrak h), \qquad A \in \Omega^1(M, \mathfrak g),
\end{equation}
where \(\mathfrak g\) and  \(\mathfrak h\) are the Lie algebras of $G$ and $H$, respectively. 

\paragraph{The gauge Lie 2-group} A  strict Lie 2-group is given by the data of a group homomorphism \(s\colon H\to G\) and an action \(\rho\colon G\times H\to H\) satisfying
\[
s(\rho(g)  h_1)=h s(h_1) h^{-1} \quad \text { and } \quad \rho (s(h_1)) h_2=h_1 h_2 h_1^{-1}
\]
for all $h_1, h_2 \in H$ and $g \in G$.

 An \emph{adjusted}  strict 2-group comes equipped  with an adjustment 
\begin{equation}
    \kappa\colon G\times \mathfrak{g} \to\mathfrak h.
\end{equation}
  The adjustment of a strict Lie 2-group must satisfy
\begin{equation}
\begin{aligned}
\kappa(s(h), x) & =h\left(\rho(x)  h^{-1}\right) \\
\kappa(g_2 g_1, x) & =\rho(g_2) \kappa(g_1, x)+\kappa(g_2, g_1 x g_1^{-1}-s(\kappa(g_1, x)))
\end{aligned}
\end{equation}
for all $g_1,g_2 \in {G}, h \in {H}$, and $x \in \mathfrak{g}$. Note that we have used the same symbol $\rho\colon \mathfrak{g} \times H \to H$ to denote  the linearization of the action $\rho\colon  G\times H \to H$. We trust the context will alleviate any ambiguity.  For examples of adjusted Lie 2-groups, one can take \(\mathcal G\) to be the loop model of the string Lie 2-algebra \cite{Saemann:2017rjm,Schmidt:2019pks,Kim:2019owc}, where both \(G\) and \(H\) are non-Abelian.

\paragraph{The adjusted  gauge Lie 2-algebra}The strict 2-group \(\mathcal G\)  linearizes to an adjusted strict Lie 2-algebra\footnote{Here $V[k]$ denotes the degree-shift of a graded vector space $V=\bigoplus_i V^i$ by $k\in\mathbb Z$, where $V[k]^i\coloneqq V^{i+k}$. Note that  $\mathfrak g$ and $\mathfrak h$ are regarded as graded vector spaces concentrated in degree zero, i.e.~$\mathfrak g=\mathfrak g^0$, so $\mathfrak h[1]\oplus\mathfrak g[0]$ has non-trivial degrees $-1$ and $0$.}  (or, equivalently, a strict two-term $L_\infty$-algebra) with graded vector space
\begin{equation}
\mathfrak G=\mathfrak h[1]\oplus\mathfrak g[0].
\end{equation} The   non-trivial graded  structure maps \footnote{Note that, by abuse of notation, we overload the symbols \(s, \rho\) and \(\kappa\) for both the finite (crossed module of Lie groups) and infinitesimal (crossed module of Lie algebras) cases; the intended meaning should be clear from the context.} on $\mathfrak G$ are
\begin{align}
    s&\colon\mathfrak h[1]\to\mathfrak g[0]&
    [-,-]&\colon\mathfrak g[0]\otimes\mathfrak g[0]\to\mathfrak g[0]\\
    \rho&\colon\mathfrak g[0]\otimes\mathfrak h[1]\to\mathfrak h[1]&
    \kappa&\colon\mathfrak g[0]\otimes\mathfrak g[0]\to\mathfrak h[1],
\end{align}
carrying degrees \(1\), \(0\), \(0\), and $-1$, respectively.  
  The  map \(\kappa\) defines the adjustment, while  $s, \rho$ and $[-,-]$ are the structure maps of  a strict  Lie 2-algebra satisfying
\[
s(\rho(x)(b)) = [x, s(b)], \qquad \rho(s(a))(b)=[a,b]
\]
for any \(x\in \mathfrak{g}\) and \(a,b\in \mathfrak{h}\).
By defining the mixed Lie bracket $[-,-]\colon \mathfrak{g}\otimes \mathfrak{h}\to \mathfrak{h}$ via the action $[x,a]\coloneqq\rho(x)(a)$ a    strict  Lie 2-algebra is equivalent to a differential graded (dg) Lie algebra concentrated in degrees $-1$ and $0$. 

Specifically, a dg~Lie algebra concentrated in degrees $-1$ and $0$ has maps 
\begin{equation}
\begin{aligned}
s\colon &&\mathfrak{h}[1]&\to \mathfrak{g}[0]\\
[-,-]\colon&&\mathfrak{g}[0]\otimes \mathfrak{h}[1]&\to \mathfrak{h}[1]\\
[-,-]\colon&&\mathfrak{g}[0]\otimes \mathfrak{g}[0]&\to \mathfrak{g}[0],
\end{aligned}
\end{equation}
where $[-,-]$ is graded-antisymmetric Lie bracket obeying the graded Jacobi identity, i.e.
\begin{align}
\left[z_1, z_2\right]&=-(-1)^{\left|z_1\right|\left|z_2\right|}\left[z_2, z_1\right],\\
\left[z_1,\left[z_2, z_3\right]\right]&=\left[\left[z_1, z_2\right], z_3\right]+(-1)^{\left|z_1\right|\left|z_2\right|}\left[z_2,\left[z_1, z_3\right]\right],
\end{align}
and $s$ is a differential graded nilquadratic derivation, i.e.
\begin{align}
s^2&=0,& s\left[z_1, z_2\right]&=\left[s z_1, z_2\right]+(-1)^{\left|z_1\right|}\left[z_1, s z_2\right].
\end{align}
In the above identities, $z_i$ are elements of homogeneous degree in $\mathfrak h[1]\oplus\mathfrak g[0]$, and $|z|$ is the degree of $z\in \mathfrak h[1]\oplus\mathfrak g[0]$.
 We shall use the dg~Lie algebra formulation in what follows.

\paragraph{Adjusted higher gauge theory} The higher gauge theory with higher gauge algebra $\mathfrak G$ has (locally) a \(\mathfrak h\)-valued two-form gauge potential \(B\in \Omega^2(M, \mathfrak{h})\)  and a  \(\mathfrak g\)-valued one-form potential \(A\in \Omega^1(M, \mathfrak{g})\), with higher gauge transformations given by
\begin{align}\label{eq:adjusted_gauge_transform}
    \delta B&= [B, \alpha]+\mathrm d\Lambda+[A,\Lambda] + \kappa(\alpha, F),\\
   \delta A&=\mathrm d\alpha + [A,\alpha] - s(\Lambda),
\end{align}
where $\alpha\in \Omega^0(M, \mathfrak{g})$,  $\Lambda\in \Omega^1(M, \mathfrak{h})$  and it is understood that the action of $s$ has been extended by linearity to $\Omega^1(M)\otimes\mathfrak h$.   In \eqref{eq:adjusted_gauge_transform},  \(H\) and \(F\) are the corresponding adjusted field strengths
\begin{align}
    H&\coloneqq\mathrm dB+[A,B]-\kappa(A, F),\\
    F&\coloneqq\mathrm dA+\frac12[A,A]+s(B).
\end{align}
Here,  $\kappa(F,A)$ is the adjustment of $H$, where for the $\mathfrak{g}$-valued forms \(A\) and \(B\) the wedge product in $\kappa(A,B)$ is understood. The role of this term is to  allow for $F\not=0$ with a consistent gauge-algebra (specifically, in the absence of an adjustment, $F\not=0$ implies the gauge-algebra does not close for $\mathfrak g$ non-Abelian). 

Under \eqref{eq:adjusted_gauge_transform}, the field strengths  transform as
\begin{align}
    \delta F&= [F,\alpha] +s(\kappa(\alpha, F)), \\
    \delta H&=[H,\alpha]+\kappa(\alpha, s(H)).
\end{align}
Note that both the 3-form \(H\) and 2-form $F$ field strengths are generically  non-Abelian. If $\kappa$ is antisymmetric, however, then $\delta H=0$ since $\kappa(s(a), x) = \rho(x)a=[x,a]$ for all $a\in \mathfrak{h}$ and $x\in \mathfrak{g}$. The Bianchi identities are then
\begin{align}
    \mathrm dF+[A,F]+s(H + \kappa(A, F) )&=0\\
    \mathrm dH+[A, H] -\kappa(A, s(H))+\kappa(F,F)&=0.
\end{align}
Note that the three-form field strength \(H\) is not closed generically and hence is not the Noether current of a magnetic \((d-4)\)-form symmetry unless \(\kappa=0\), as in the case of Abelian 2-form electrodynamics.\footnote{If, however, $\kappa$ is antisymmetric, then  $\delta H=0$ and $\mathrm dH=0$, even if $\mathfrak{h}$ is non-Abelian; in that case, \(H\) is the Noether current for a \((d-4)\)-form symmetry.}

Finally,  assuming invariant  inner products $\langle-,-\rangle_{\mathfrak h}$  and $\langle-,-\rangle_{\mathfrak g}$ on \(\mathfrak h\) and \(\mathfrak g\) respectively, we may straightforwardly write down the higher gauge theory action
\begin{equation}
    S_0[A,B] \coloneqq \int\frac12\langle H, \star H\rangle_{\mathfrak h}+\frac12\langle F, \star F\rangle_{\mathfrak g}, 
\end{equation}
where the wedge product between the forms $A$ and $B$ in $\langle A, B\rangle$ is understood. 

\subsection{Higher-form symmetries of adjusted higher gauge theory}

Here we consider the global higher $p$-form symmetries of the adjusted higher form gauge theories discussed above. 
In particular,  we shall  determine the adjusted higher gauge theory analogue of the well-known Yang--Mills    one-form centre symmetry.

The adjusted higher gauge theory  $p$-form centre symmetry is most readily identified in analogy to the classic derivation of the Yang--Mills centre symmetry, which we briefly recall here to illustrate the parallel.
Consider a Euclidean spacetime \(M=\Sigma\times\mathbb S^1\) with periodic time \(t\sim t+\beta\), where \(\beta\) may be interpreted as inverse temperature.
In the classical case of Euclidean Yang--Mills theory with gauge group $G$, if we for simplicity work with the trivial \(G\)-bundle on \(M\), then the potential \(A\) is periodic, i.e.\ \(A(\vec x,t+\beta)=A(\vec x,t)\). The potential transforms under a gauge transformation \(g\colon M\to G\) as
\begin{equation}\label{eq:YM_gauge_transformation}
    A\mapsto g^{-1} Ag + g^{-1}\mathrm d g.
\end{equation}
Since \(g\) is periodic (being defined on \(M=\Sigma\times\mathbb S^1\)), \(A\) remains periodic after gauge transformation.
One then considers non-periodic gauge-like transformations formally given by \eqref{eq:YM_gauge_transformation} but now parameterized by a not-necessarily-periodic \(g\colon \Sigma\times\mathbb R \to G\) satisfying the $G$-twisted periodicity condition
\begin{align}\label{eq:non-periodic_gauge}
    g(\vec x, t +\beta) &= \gamma g(\vec x, t),\\
    g(\vec x,t+\beta)^{-1}\mathrm dg(\vec x,t+\beta)
    &=g(\vec x,t)^{-1}\mathrm dg(\vec x,t)
\end{align}
for some \(\gamma\in G\).
For this to be a symmetry of the theory,
the periodicity of the gauge potential, $A(\vec x, t +\beta) = A(\vec x, t)$,  must be preserved under \eqref{eq:YM_gauge_transformation}, that is, we require
\begin{multline}
    g^{-1}(t+\beta)A(t+\beta)g(t+\beta)+g(t+\beta)^{-1}\mathrm dg(t+\beta)\\
    =
    g^{-1}(t)A(t)g(t)+g(t)^{-1}\mathrm dg(t),
\end{multline}
where we have suppressed the dependence on \(\vec x\) for clarity.
Since \(A\) and \(g^{-1}\mathrm dg\) are periodic, this reduces to
\begin{equation}
    \gamma^{-1}g^{-1}(t)A(t)g(t)\gamma
    =
    g^{-1}(t)A(t)g(t).
\end{equation}
Since the above must hold for arbitrary \(g^{-1}(t)A(t)g(t)\in G\), the group element \(\gamma\) must belong to the centre \(\operatorname Z(G)\) of the gauge group (e.g.\ \(\operatorname Z(\operatorname{SU}(N))=\mathbb Z_N\) when \(G=\operatorname{SU}(N)\)).
Note that, while taking the form of a gauge transformation and leaving the Yang--Mills action invariant,  \eqref{eq:non-periodic_gauge} has a non-trivial action on certain gauge-invariant observables such as  Polyakov loops and so has physical implications (true gauge transformations must be periodic, not twisted-periodic as in \eqref{eq:non-periodic_gauge}).
This implies that pure Yang--Mills theory has a \(\operatorname Z(G)\)-valued one-form symmetry; in particular, that the centre symmetry is a one-form symmetry is apparent from the fact that Wilson lines in the fundamental representation transform under $\gamma$, with $\operatorname Z(G)$-valued charge (see e.g.~\cite{Gomes:2023ahz}).
Furthermore, the symmetry continues to exist if there are additional adjoint-valued vector fields but generically breaks if there exists matter transforming under other (e.g.\ fundamental) representations.

A similar analysis holds for adjusted higher gauge theory. To illustrate this analogy  we need the \emph{finite} adjusted  higher gauge transformations, which are given by 
\begin{equation}\label{eq:adjusted_gauge_transform_finite}
\begin{aligned}
A & {\overset{(g,\Lambda)}\mapsto} g^{-1} A g+g^{-1}\mathrm dg-s(\Lambda) \\
B & {\overset{(g,\Lambda)}\mapsto} \rho(g^{-1})B +\mathrm{d} \Lambda+\left[g^{-1} A g+g^{-1}\mathrm dg-s(\Lambda),  \Lambda\right]\\
&\qquad\qquad+\frac{1}{2}\left[\Lambda, \Lambda\right]-\kappa\mleft(g^{-1}, F\mright),
\end{aligned}
\end{equation}
where the gauge parameters are \(g\colon M\to G\) and \(\Lambda \in \Omega^1(M,\mathfrak h)\)
(see \cite[(3.35)]{Borsten:2024gox} and references therein for  further details).
Similarly to Yang--Mills theory, identifying the (higher-form) centre symmetry amounts to determining global higher gauge transformations \((g,\Lambda)\) in \eqref{eq:adjusted_gauge_transform_finite} that leave the potentials \(A\) and \(B\) periodic (assuming, again for simplicity, a trivial \(\mathcal G\)-bundle on spacetime \(M=\Sigma\times\mathbb S^1\)):
\begin{equation}\label{eq:non-periodic_higher_gauge}
\begin{aligned}
    g(\vec x,t+\beta)&=\gamma g(\vec x,t),\\
    g(\vec x,t+\beta)^{-1}\mathrm dg(\vec x,t+\beta)&=g(\vec x,t)^{-1}\mathrm dg(\vec x,t),\\
    \Lambda(\vec x,t+\beta)&=\Lambda(\vec x,t)+\lambda,\\
    \mathrm d\Lambda(\vec x,t+\beta)&=\Lambda(\vec x,t)
\end{aligned}
\end{equation}
for \(\gamma\in G\) and \(\lambda\in\mathfrak h\otimes\mathrm T^*_{(\vec x,t)}M\).
From  \eqref{eq:adjusted_gauge_transform_finite}, we see that gauge transformations \((g,\Lambda)\) leave the potentials \(A\) and \(B\) periodic if \(\lambda\in (\ker s\cap\operatorname Z(\mathfrak h))\otimes\mathrm T^*_{(\vec x,t)}M\) and \(\gamma\in\operatorname Z(G)\) and
\begin{align}
    \kappa(\gamma^{-1}, x)&=0\;\forall x\in\mathfrak g,&
    \rho(\gamma)&=\operatorname{id}_{\mathfrak h},
\end{align}
where, again overloading the notation, \(\rho\colon G\times \mathfrak h \to \mathfrak h\) is  the induced action of \(G\) on \(\mathfrak h\). Thus, the 2-group centre symmetries are given by the 2-group described by the crossed module
\begin{multline}
    \operatorname Z(\mathcal G) = \Big((\ker s)\cap\operatorname Z(\mathfrak h) \to \\\left\{\gamma\in\operatorname Z(G)|\kappa(\gamma^{-1},\mathfrak g)=0,\;\rho(\gamma)=\operatorname{id}_{\mathfrak h}\right\}\Big)
\end{multline}
with trivial structure maps. Note that this 2-group is Abelian in the strongest possible sense (i.e.\ all structure maps are trivial) in agreement with the expectation that higher-form symmetries must be Abelian due to the Eckmann--Hilton argument.
For lack of a better name, we call this the \emph{adjusted centre symmetry 2-group} of $\mathcal{G}$ \footnote{
    In fact, since \(\gamma\) is valued in the group \(\operatorname Z(G)\subset G\) whereas \(\lambda\) is valued in \(\operatorname Z(\mathfrak h)\subset\mathfrak h\), this is better regarded as the centre of a differential graded Lie group with body \(G\) and with nonzero-degree coordinates corresponding to \(\mathfrak h\) obtained as a partial linearization of the 2-group \(\mathcal G\).
}. Note, although there is the familiar $Z(G)$ factor, $\operatorname Z(\mathcal G)$ has more structure than the typical instances of discrete $p$-form symmetries and arises from a non-Abelian higher gauge theory directly. 

Now, just as the centre symmetry group of Yang--Mills theory is in fact a one-form symmetry valued in a (1-)group, the  centre symmetry 2-group of our higher gauge theory is a  pair of one-form and two-form symmetries that combine into a 2-group. 

The symmetry topological defects generating the $2$-form center symmetry and the corresponding charged branes are a direct generalisation of the 1-form centre symmetry for Yang--Mills theory. In the latter case, symmetry topological defects are  co-dimension-2 Gukov--Witten operators that can be regarded  equivalently  as  \eqref{eq:YM_gauge_transformation} and \eqref{eq:non-periodic_gauge} for $\gamma\in Z(G)$ in the conventional centre symmetry language or as a Wilson line operator of a $\mathbb{Z}_N$-valued background 1-connection $A_{\mathbb{Z}_N}$ (see e.g.~\cite{Brennan:2023mmt}). The charged objects are the conventional Wilson line operators. 
	
	Analogously, for the 2-group higher gauge theory symmetry, the  topological defects are the co-dimension-3 Gukov--Witten-type operators that can be regarded  equivalently  as \eqref{eq:adjusted_gauge_transform_finite}  and \eqref{eq:non-periodic_higher_gauge} for $(\lambda, \gamma)\in Z(\mathcal G)$ in the conventional centre symmetry language or a Wilson surface operator of $\mathbb{Z}_N$-valued background 2-connection, locally $(A_{\mathbb{Z}_N}, B_{\mathbb{Z}_N}$).  The charged objects are the  Wilson surface  operators taking values in the 2-group $\mathcal{G}$, given by the path-ordered exponentials of  $B$  integrated over a surface and $A$ integrated over the associated  boundary in a coherent fashion. More precisely, Wilson surface  operators are  2-holonomies of the 2-connection (i.e.~locally $B$ and $A$), which are  equivalent to the 2-functors from the 2-path groupoid on $M$ to the 2-group $\mathcal{G}$, as described in \cite{Baez:2004in} but where one has to account for the failure of fake flatness using the adjustment, as described in \cite{Kim:2019owc}.

In principle \footnote{In practice, however, it quickly becomes impractical to spell out the $p$-group coherence relations as $p$ grows.}, this entire discussion readily generalizes to arbitrary Lie $p$-group higher gauge theories, with corresponding $p$-group generalized global symmetry.  Here, we simply assume that the highest component  is a $\mathbb Z_N$ $p$-form symmetry, which we take to be our microscopic global symmetry in order to generalize the results of \cite{Nguyen:2024ikq}. 

\subsection{Coupling to adjoint matter: higher gauge Coulomb phases}\label{sec:highermatter}
Let us couple the gauge potentials \(A\) and \(B\) to matter transforming in the adjoint representation of \(\mathcal G\), formally given by the associated 2-bundle. Thus, we have the fields
\begin{align}
    \phi\in\Omega^1(M;\mathfrak h), \qquad
    \chi\in\Omega^0(M;\mathfrak g),
\end{align}
with associated covariant derivatives modelled after the field strengths:
\begin{align}
\mathrm D\chi&=\mathrm d\chi+[A,\chi]+s(\phi) \\
      \mathrm D\phi&=\mathrm d\phi+[A,\phi]+[\chi,B]+\kappa(F,\phi)+\kappa(\mathrm D\chi,A).
\end{align}

Similarly, the higher gauge transformations are modelled after the potentials,
\begin{align}\label{eq:adjusted_matter_transform}
  \delta  \phi&= [\alpha, \phi]+[\chi,\Lambda] +  \kappa(\alpha, \mathrm D \chi),\\
 \delta   \chi&= [\chi,\alpha],
\end{align}
so that the covariant derivatives transform in the same manner as the field strengths
\begin{align}
\delta \mathrm D\phi&=[\mathrm D \phi,\alpha]+\kappa(\alpha, s(\mathrm D \phi)),\\
\delta \mathrm D\chi&=[\mathrm D \chi,\alpha] +s(\kappa(\alpha, \mathrm D \chi)). 
\end{align}

Thus, we may immediately deduce that  the invariant minimally-coupled  action is given by 
\begin{equation}
    S[A, B, \phi, \chi] = S_0[A,B] + \int \frac12\langle \mathrm D\phi, \star \mathrm D\phi \rangle_{\mathfrak{g}}+\frac12\langle\mathrm D\chi, \star\mathrm D\chi \rangle_{\mathfrak{h}}.
\end{equation}

Now, assume that the scalar field \(\chi\) obtains a vacuum expectation value \(\chi_0\) of order \(m\), via spontaneous symmetry breaking upon including a potential term. In that case, we obtain mass terms for \(A\) and \(B\) of the form
\[
     \int\frac12\left\langle [A,\chi_0],\star[A,\chi_0]\right\rangle
     +\frac12\left\langle[B,\chi_0],\star[B,\chi_0]\right\rangle,
\]
so that only the subalgebra left invariant under the action by \(\chi_0\) will remain massless; assuming semisimplicity and generic values of \(\chi_0\), this means that the surviving subalgebras of \(\mathfrak g\) and \(\mathfrak h\) will be Cartan subalgebras, so the massless remnants of \(A\) and \(B\) will be purely Abelian.
In this case, the global higher-form (centre) symmetry survives intact and will also be visible from the low-energy effective theory.  We now  turn to an analysis of the other possible  phases  taking a bottom-up approach, starting with the expected low energy effective  $BF$ model.

\section{Effective deformed \emph{p}-form \emph{BF} models}\label{sec:bf}

In the previous section we considered purely adjoint matter and argued that, generically, this leads to a higher analogue of the Coulomb phase. Now suppose, instead,  that we couple the system to matter fields in some arbitrary fashion (that is, with matter fields not in the  adjoint representation). In this case   the matter condenses and  we, generically, break the entirety of the gauge symmetry \(\mathcal G\). This process generally yields a Higgs phase described by an effective \(BF\) model, for example as witnessed in the case of Higgsed Maxwell theory \cite{Banks:2010zn} (reviewed in e.g.~\cite{Brennan:2023mmt,Bhardwaj:2023kri}).

Motivated by this general picture, in this section we take a bottom-up approach supposing that microscopically we have $\mathbb Z_N$ $p$-form symmetry which is broken, resulting in an effective  $p$-form $BF$ model. However, as described below (see also \cite{Nguyen:2024ikq}), the $p$-form $BF$  model has more than  the required $\mathbb Z_N$ $p$-form symmetry. Consequently, the extra symmetry must be broken by certain class-preserving deformations.  For ordinary and higher effective $BF$ models alike,  partially breaking  the global $p$-form symmetries for $d>3$, as in \cite{Nguyen:2024ikq}, requires a deformation given by a minimal coupling of scalar fields to the higher $p$-potentials. If $p>1$  then obviously the minimal coupling given by the  naive covariant derivative is not possible.  There are various possible solutions, in particular minimally coupling to higher matter fields as in \autoref{sec:highermatter}. Here instead, we shall use the notion of Chen forms \cite{zbMATH03361026,zbMATH03532253,zbMATH04050538,Baez:2004in,getzlerjonespetrack} (also called \emph{iterated integrals}; for an informal introduction, see \cite[App.~H]{Kim:2019owc}).  to minimally  couple $p$-forms to ordinary (zero-form) scalar matter. For simplicity, throughout the following we assume  \(d-p-2>0\),  time reversal symmetry (this forbids e.g.\ Chern--Simons topological mass terms), and that the homology classes of spacetime \(M\) are finitely generated and torsion-free.

\subsection{Chen forms}

In \cite{Nguyen:2024ikq}, the magnetic  $(d-2)$-form symmetry exhibited by the effective one-form $BF$ model is explicitly broken by minimally coupling a scalar via $\mathrm D \phi = (\mathrm d + B)\phi$.  This is required by the assumption that the microscopic theory has only a $\mathbb Z_N$ one-form symmetry.  However, this is only straightforwardly possible in $d=3$, where $B$ is a one-form.  To generalize this construction to $p$ and $d$ for which $d-p-1\not=1$, we employ the Chen form construction.

Given a smooth manifold \(M\), the (unbased) loop space \(\mathcal LM\) is the space of smooth functions \(\mathbb S^1\to M\); it is an infinite-dimensional manifold. Let \(A\) be a \((p+1)\)-form on \(M\). Then  the associated \emph{Chen form}
 is a \(p\)-form \(\check A\) on \(\mathcal LM\) defined as
\begin{equation}
    \check A|_\gamma = \int_{\mathbb S^1}\mathrm dt\,\frac{\partial\gamma}{\partial t}\mathbin\lrcorner\operatorname{ev}^*_tA,
\end{equation}
where \(\operatorname{ev}^*_t\) is the pullback of a differential form along the evaluation map
\begin{equation}
\begin{aligned}
    \operatorname{ev}_t\colon\mathcal LM&\to M\\
    \gamma&\mapsto\gamma(t),
\end{aligned}
\end{equation}
and \(\mathbin\lrcorner\) is the interior product with the vector field \(\partial\gamma/\partial t\) defined on \(\mathcal LM\).
Iterating this construction, one obtains from a \((p+k)\)-form \(A\in\Omega^{p+k}(M)\) a \(p\)-form \(\check A\in\Omega^p(\mathcal L^k(M))\).

The loop space and Chen forms are central to string field theory. A particle living on spacetime \(M\) is given by a worldline \(\mathbb R\to M\), and second quantization yields a scalar field living on \(M\). A string living on a spacetime \(M\) is given by a worldsheet \(\mathbb R\times\mathbb S^1\to M\), which is equivalent to a map \(\mathbb R\to\mathcal LM\), that is, a worldline in the loop space of spacetime; a second quantization yields a scalar field living on \(\mathcal LM\), which is in turn equivalent to an infinite tower of fields living on \(M\) by Fourier-series expansion. More generally, a toroidal \(k\)-brane living on a spacetime \(M\) is given by a field living on \(\mathcal L^kM\).

\subsection{Phases of the deformed $p$-form \(BF\) model}
We shall assume that in  the deep infrared, on a spacetime manifold \(M\), the low-energy effective action of a theory with a microscopic \(\mathbb Z_N\) \(p\)-form symmetry should be given by a \(BF\) model \cite{Brennan:2023mmt,Bhardwaj:2023kri}
\begin{equation}
    S_0 = \int_M\frac{\mathrm iN}{2\pi}B\wedge\mathrm dA,
\end{equation}
where \(A\) is a \(p\)-form and \(B\) is a \((d-p-1)\)-form and the  gauge transformations are given by 
\begin{align}
    A&\mapsto A+\mathrm dc&
    B&\mapsto B+\mathrm d\Lambda
\end{align}
for \(c\in\Omega^{p-1}(M)\) and \(\Lambda\in\Omega^{d-p-2}(M)\). 

Let us briefly summarise the discrete $p$-form symmetries. See \cite{Brennan:2023mmt,Bhardwaj:2023kri} for  detailed treatments of $d$-dimensional (continuous and discrete) $p$-form $BF$ theory and its higher global symmetries. 
This model has a \(\mathbb Z_N\)-valued \(p\)-form symmetry as well as a \(\mathbb Z_N\)-valued \((d-p-1)\)-form symmetry since \(A\) and \(B\) are on an equal footing;
that is, the global internal symmetry group is the Abelian \(\infty\)-group
\begin{equation}\label{eq:bf_symmetry}
    (\mathbb Z_N)[-p]\oplus(\mathbb Z_N)[1+p-d].
\end{equation}
The electric $p$-form symmetry $\mathbb Z_N[-p]$  is generated by the ``current'' 
\begin{equation}
\star  J_\mathrm{el}=\frac{N B}{2 \pi},
\end{equation}
and  the usual Wilson operator for $B$ is the symmetry defect operator, which acts on the Wilson operators for $A$ (and vice versus).

In generalizing \cite{Nguyen:2024ikq}, our basic assumption is that the microscopic symmetry is only the \(\mathbb Z_N\)-valued \(p\)-form symmetry. Thus, the $p$-form $BF$ action has too much symmetry and, again following \cite{Nguyen:2024ikq}, we explicitly break the \((d-p-1)\)-form symmetry by deforming the action. Specifically, we  break the extra symmetry by coupling \(B\) to a \((d-p-2)\)-brane with unit charge. This brane can be represented by a complex scalar field \(\phi\) living on the iterated loop space \(\mathcal L^{d-p-2}M\), so that we obtain
\begin{equation}
    S_1 =
    \int_{\mathcal L^{d-p-2}M}(d+\mathrm i\check B)\phi\wedge\star((d-\mathrm i\check B)\phi^*).
\end{equation}
where \(\check B\) is a one-form living on \(\mathcal L^{d-p-2}M\) obtained by the Chen-form construction from \(B\).

Having coupled a source to \(B\), its equation of motion no longer enforces \(\mathrm dA=0\), and  \(\mathrm dA\) becomes the Noether current for a \(\operatorname U(1)\)-valued \((d-p-2)\)-form symmetry. We may break this symmetry as well by introducing a kinetic term for \(A\) and coupling to magnetic currents:
\begin{equation}
    S_2 = \int\frac1{2g_A^2}\mathrm dA\wedge\star\mathrm dA + \tilde A\wedge J,
\end{equation}
where \(\mathrm d\tilde A=\star\mathrm dA\) and \(J\) is a \((p+2)\)-form operator built out of matter fields.

Given the action \(S=S_0+S_1+S_2\), let us consider the various phases.
\begin{itemize}
\item If we additionally have mass terms or strongly coupled interactions that render the fields \(\phi\) and \(J\) heavy at a mass scale \(M\), then in the deep infrared \(\ll M\), \(S_1\) and \(S_2\) become irrelevant, and we end up with the action \(S_0\) describing the Higgs phase, where the symmetry is enhanced from \(\mathbb Z_N[-1]\) to \eqref{eq:bf_symmetry}.
\item Suppose that we additionally have mass terms for \(J\) so that at low energy scales \(J\) can be integrated out, but that a component of the string field \(\phi\) condenses and acquires vacuum expectation values.
In that case, \(B\) becomes massive and may be integrated out, and similarly for \(J\). The low-energy action then becomes
\[
    S = \int\frac1{2g^2_A}\mathrm dA\wedge\star\mathrm dA+\dotsb,
\]
with no monopole operators for \(A\). This is a Coulomb phase with a low-energy \(\operatorname U(1)\) gauge field \(A\).
\item If both \(\phi\) and \(J\) acquire vacuum expectation values, then we obtain a \(\operatorname U(1)\) gauge theory perturbed by monopole operators:
\[
    S = \int\frac1{2g^2_A}\mathrm dA\wedge\star\mathrm dA+\tilde A\wedge J+\dotsb.
\]
The term \(\tilde A\wedge J\) may be relevant \cite{Polyakov:1975rs,Polyakov:1976fu} and lead to unbroken \(\mathbb Z_N\) one-form symmetry.
\end{itemize}
Note that, in the case \(p+2=d\), then one can generically take \(J\propto \phi^N+\text{c.c.}\), as noted in \cite{Nguyen:2024ikq}. In such cases, if one cannot independently set \(\phi\) and \(J\), one might not have a Coulomb phase.

Here we deformed the $BF$ theory by introducing a $(d-p-2)$-brane with unit charge. It would be interesting to understand if this can be related to cocycle twist
deformations  appearing in various aspects of topological quantum field theories such as Dijkgraaf-Witten theory (themselves often related to $BF$ theories),  and generalized symmetries, e.g.~\cite{Dijkgraaf:1989pz,Putrov:2016qdo,Guo:2018vij,Delcamp:2019fdp}.  

\section{Coulomb phases in  lattice \emph{p}-form electrodynamics}\label{sec:lattice}
We illustrate the claim with a lattice model, namely a modified Villain model, where the mechanism of the Coulomb phase is particularly transparent. This is a straightforward higher analogue of the modified Villain model studied in \cite{Nguyen:2024ikq}; see also \cite{Orland:1981ku} for an earlier discussion of higher-form \(\operatorname U(1)\) gauge theory in the Villain form.

\subsection{Cellular cochains as lattice analogues of differential forms}
In a lattice model, the lattice structure equips spacetime with the structure of a cellular complex (or CW-complex), so that it is natural to use the language of cellular cohomology, such that cellular cocycles may be thought of as discrete analogues of differential forms. (For a pedagogical description of cellular homology, see \cite{hatcher}.)
Let us consider a tiling (or tessellation)\footnote{We do not need periodicity; an aperiodic tiling will also do.} \(\Lambda\) of \(\mathbb R^d\) --- for instance, by hypercubes, or the triangular or hexagonal tiling of \(\mathbb R^2\), or the tiling of \(\mathbb R^3\) by truncated octahedra or more complicated polyhedra. In technical terms, this defines a cell complex structure on \(\mathbb R^d\). We use the notation \(X_k(\Lambda)\) for the \(k\)-cells of this cell complex. Thus \(X_0(\Lambda)\) is the set of vertices; \(X_1(\Lambda)\) is the set of links (edges); \(X_2(\Lambda)\) is the set of plaquettes. For instance, when \(\Lambda\) is the hypercube tiling, we have
\begin{align}
    X_0(\Lambda)&=\left\{\{v\}\middle|v\in\mathbb Z^d\right\}\\
    X_1(\Lambda)&=\left\{\{v+su|s\in[0,1]\}\middle|v\in\mathbb Z^d,\;u\in\{\mathbf e_1,\dotsc,\mathbf e_n\}\right\}\\
    X_2(\Lambda)&=\Big\{\{v+s_1u_1+s_2u_2|s_1,s_2\in[0,1]\}\Big|v\in\mathbb Z^d,\\
    &\qquad\qquad\qquad\qquad\qquad u_1\in\{\mathbf e_1,\dotsc,\mathbf e_n\}\ni u_2\Big\},\notag
\end{align}
where \((\mathbf e_1,\dotsc,\mathbf e_n)\) are the canonical basis vectors of \(\mathbb R^n\).
Tilings admit a duality, generalizing that of planar graphs, which may be thought of as the `infinite-volume limit' of the duality of convex polytopes. If \(\hat\Lambda\) is the tiling dual to \(\Lambda\), then we have the canonical identifications
\begin{equation}\label{eq:lattice-duality}
    X_k(\Lambda)=X_{d-k}(\hat\Lambda).
\end{equation}
For instance, the triangular tiling of \(\mathbb R^2\) is dual to the hexagonal tiling, and the hypercube tiling of \(\mathbb R^d\) is dual to itself.
On such a tiling, we use the notation \(\Omega^k(\Lambda;G)\) (where \(G\) is an Abelian group under addition) for the set of functions
\begin{equation}
    X_k(\Lambda)\to G.
\end{equation}
When \(G=\mathbb R\), this should be thought of as a discrete analogue of the space of differential \(k\)-forms on \(\mathbb R^d\); hence we denote the function evaluation as \(\alpha|_x\) instead of \(\alpha(x)\). On the other hand, in the discrete context now it makes sense to talk of \(\mathbb Z\)-valued \(k\)-forms, for instance. 

Let \(C_k(\Lambda)\) be the free Abelian group of \(\mathbb Z\)-linear combinations of elements of \(X_k(\Lambda)\). (These are the chains in the cellular homology associated to the cell complex \(\Lambda\).) Thus, \(\Omega^k(\Lambda;G)\) may be equivalently thought of as the set of Abelian group homomorphisms (i.e.\ \(\mathbb Z\)-linear maps)
\begin{equation}
    C_k(\Lambda)\to G.
\end{equation}
Mathematically, \(\Omega^k(\Lambda;G)\) is the set of \(G\)-valued cellular cochains.
On \(C_k(\Lambda)\), there exists a standard notion of boundary, where we keep track of the orientation,
\begin{equation}
    \partial\colon C_k(\Lambda)\to C_{k-1}(\Lambda),
\end{equation}
which squares to zero, i.e.\ \(\partial^2=0\); this defines cellular homology \cite{hatcher}.

While we do not define the wedge product of the `differential forms' on \(\Lambda\), we may define their exterior derivative as the Wilson hypersurfaces around each cell. That is, suppose that the boundary of a cell \(x\) is \(\partial x=l_1+l_2+\dotsb+l_m\). For \(\alpha\in\Omega^k(\Lambda;G)\), then \(\mathrm d\alpha\in\Omega^{k+1}(\Lambda;G)\) is defined as
\begin{equation}
    \mathrm d\alpha|_x = \alpha|_{l_1}+\alpha|_{l_2}+\dotsb+\alpha|_{l_m}
\end{equation}
for \(x\in C_{k+1}(\Lambda)\). This defines a map
\begin{equation}
    \mathrm d\colon\Omega^p(\Lambda;G) \to \Omega^{p+1}(\Lambda;G).
\end{equation}
This operator squares to zero; the corresponding cohomology is cellular cohomology with coefficients in \(G\). Finally, there also exists an analogue of the Hodge dual operator
\begin{equation}
    \star\colon\Omega^k(\Lambda;G)\to\Omega^{d-k}(\hat\Lambda;G)
\end{equation}
defined by means of the isomorphism \eqref{eq:lattice-duality}.

\subsection{Villain model of \(p\)-form electrodynamics}
In a Villain model (first introduced in \cite{Villain:1974ir}), one resolves the group \(\operatorname U(1)\cong\mathbb R/\mathbb Z\) by considering an \(\mathbb R\)-valued gauge field and then subsequently gauging a \(\mathbb Z\)-valued two-form symmetry. Analogously, one may realize the gauge group \(\mathbb Z_N\cong N^{-1}\mathbb Z/\mathbb Z\) by starting with a \(N^{-1}\mathbb Z\)-valued gauge field and then gauging an \(\mathbb Z\)-valued two-form symmetry;
similarly, a \(\mathbb Z_N\)-valued \(p\)-form symmetry may be obtained by starting with an \(N^{-1}\mathbb Z\)-valued \(p\)-form symmetry whose \(\mathbb Z\)-valued subgroup is then gauged away.

Suppose that we are given a tiling \(\Lambda\) of \(\mathbb R^d\) with associated dual tiling \(\hat\Lambda\). A Villain model of a \(p\)-form electrodynamics with a discrete gauge group \(\mathbb Z_N\cong N^{-1}\mathbb Z/\mathbb Z\) is given by the following action:
\begin{equation}
\begin{aligned}
    S(a,m;\beta) &= \frac12\beta\sum_{X_{p+1}(\Lambda)}(\mathrm da+m)^2 \\
    &= \frac12\beta\sum_{x\in X_{p+1}(\Lambda)}\left((\mathrm da)|_x+m|_x\right)^2
\end{aligned}
\end{equation}
for dynamical fields
\begin{align}
    a&\in\Omega^p(\Lambda;N^{-1}\mathbb Z)&
    m&\in\Omega^{p+1}(\Lambda;\mathbb Z),
\end{align}
and where \(\beta\in\mathbb R\) is a coupling constant playing the role of inverse temperature.
As explained in \cite{Nguyen:2024ikq}, this model now has two different kinds of defects: in addition to monopoles for \(a\) (which are \((d-p-2)\)-branes), one also has centre vortices (analogues of the QCD centre vortices described in \cite{tHooft:1977nqb,Cornwall:1979hz,Nielsen:1979xu}), which are \((d-p-1)\)-branes.
Each monopole is attached to \(N\) centre vortices.

Given the bestiary of defects, there are three possible phases:\footnote{We neglect such complications as when some but not all of centre vortices condense, corresponding to a breaking of \(\mathbb Z_N\) into a subgroup; see \cite[App.~E]{Nguyen:2024ikq} for a detailed discussion.}
\begin{itemize}
\item Higgs phase, when both monopoles and centre vortices are heavy (\(\beta\to\infty\)).
\item Confined phase, when both monopoles and centre vortices are light (\(\beta\to0\)).
\item Coulomb phase, when monopoles are heavy but centre vortices are light.
\end{itemize}
The case where monopoles are light but centre vortices are heavy is not possible since each monopole is attached to \(N\) centre vortices. Now, we claim that, in the Coulomb phase as defined above, there is a light \(\operatorname U(1)\) \(q\)-form field for \(q=d-p-2\), as the name suggests.

The argument for the existence of the Coulomb phase is as follows. To see it, we must first suppress the monopoles. We may force this by introducing a Lagrange multiplier \(\tilde a\in\Omega^{d-p-2}(\hat\Lambda;\mathbb R)\) to kill the monopole operator \(\mathrm dm\); the resulting model is the modified Villain action \cite{Sulejmanpasic:2019ytl,Gorantla:2021svj}
\begin{equation}
    S_\mathrm{mV}(a,\tilde a,m;\beta) \coloneqq \frac12\beta\sum_{X_{p+1}(\Lambda)}(\mathrm da+m)^2
    -\mathrm i\sum_{X_{p+1}(\Lambda)}m(\star\mathrm d\tilde a).
\end{equation}
Now, a Poisson resummation yields
\begin{multline}
    \sum_{m\in\Omega^{p+1}(\Lambda)}\exp(-S_\mathrm{mV}(a,\tilde a,m;\beta))
    =\\
    \sum_{\tilde m\in\Omega^{p+1}(\Lambda)}\exp(-\tilde S_\mathrm{mV}(a,\tilde a,\tilde m;\tilde\beta))
\end{multline}
where $\tilde\beta =\beta^{-1}$ and the dual action is defined as
\begin{multline}
    \tilde S_\mathrm{mV}(a,\tilde a,\tilde m;\tilde\beta)
    \coloneqq\\\frac12\tilde\beta\sum_{X_{d-p-1}(\hat\Lambda)}(\mathrm d\tilde a+\star\tilde m)^2
    +\mathrm i\sum_{X_{p+1}(\Lambda)}\tilde m(\star\mathrm da).
\end{multline}
Consequently, the dual model describes a \(\operatorname U(1)\)-valued \((d-p-2)\)-form gauge field in Villain form, but with the Lagrange multiplier \(a\) constraining \(\mathrm d\tilde m\). But \(a\) is quantized in units of \(N^{-1}\mathbb Z\), which means that \(\mathrm d\tilde m\) only vanishes modulo \(N\). So, we get a \(\operatorname U(1)\) gauge theory coupled to magnetic monopoles of charge \(N\). Now, when \(\beta\to0\) (high temperature limit) in the original \(S_\mathrm{mV}\), so that centre vortices proliferate but monopoles are still suppressed, then we see that we get a light U(1) gauge field, so this is a Coulomb phase.

In the above, we have simply completely suppressed the magnetic monopoles using a Lagrange multiplier. If instead one simply gives the magnetic monopoles a very large but finite mass \(M\gg1\) using a term like
\begin{equation}
    \frac12M^2\sum_{X_{p+2}(\Lambda)}(\mathrm dm)^2,
\end{equation}
one expect the qualitative picture to be the same, following symmetry-based arguments as in \cite{Nguyen:2024ikq}.

\section*{Acknowledgements}
The authors thank Tin Sulejmanpašić for helpful discussion.

\bibliography{biblio}

@article{Guo:2018vij,
    author = "Guo, Meng and Ohmori, Kantaro and Putrov, Pavel and Wan, Zheyan and Wang, Juven",
    title = "{Fermionic Finite-Group Gauge Theories and Interacting Symmetric/Crystalline Orders via Cobordisms}",
    eprint = "1812.11959",
    archivePrefix = "arXiv",
    primaryClass = "hep-th",
    doi = "10.1007/s00220-019-03671-6",
    journal = "Commun. Math. Phys.",
    volume = "376",
    number = "2",
    pages = "1073--1154",
    year = "2020"
}

@article{Putrov:2016qdo,
    author = "Putrov, Pavel and Wang, Juven and Yau, Shing-Tung",
    title = "{Braiding Statistics and Link Invariants of Bosonic/Fermionic Topological Quantum Matter in 2+1 and 3+1 dimensions}",
    eprint = "1612.09298",
    archivePrefix = "arXiv",
    primaryClass = "cond-mat.str-el",
    doi = "10.1016/j.aop.2017.06.019",
    journal = "Annals Phys.",
    volume = "384",
    pages = "254--287",
    year = "2017"
}

@article{Dijkgraaf:1989pz,
    author = "Dijkgraaf, Robbert and Witten, Edward",
    title = "{Topological Gauge Theories and Group Cohomology}",
    reportNumber = "THU-89-9, IASSNS-HEP-89-33",
    doi = "10.1007/BF02096988",
    journal = "Commun. Math. Phys.",
    volume = "129",
    pages = "393",
    year = "1990"
}

@article{Delcamp:2019fdp,
    author = "Delcamp, Clement and Tiwari, Apoorv",
    title = "{On 2-form gauge models of topological phases}",
    eprint = "1901.02249",
    archivePrefix = "arXiv",
    primaryClass = "hep-th",
    doi = "10.1007/JHEP05(2019)064",
    journal = "JHEP",
    volume = "05",
    pages = "064",
    year = "2019"
}

@article{Kovner:1992pu,
    author = "Kovner, Alexander and Rosenstein, Baruch",
    title = "New look at {QED}\textsubscript4: The Photon as a {G}oldstone boson and the topological interpretation of electric charge",
    eprint = "hep-th/9210154",
    archivePrefix = "arXiv",
    reportNumber = "LA-UR-92-3478",
    doi = "10.1103/PhysRevD.49.5571",
    journal = "Physical Review~D",
    volume = "49",
    pages = "5571--5581",
    year = "1994",
    month = may,
    number = 10,
}

@article{Hellerman:2006zs,
    author = "Hellerman, Simeon and Henriques, André Gil and Pantev, Tony G. and Eric Redmund Sharpe and Ando, Matthew",
    title = "Cluster decomposition, {T}-duality, and gerby {CFT}s",
    eprint = "hep-th/0606034",
    archivePrefix = "arXiv",
    doi = "10.4310/ATMP.2007.v11.n5.a2",
    journal = "Advances in Theoretical and Mathematical Physics",
    volume = "11",
    number = "5",
    pages = "751--818",
    year = "2007"
}

@article{Pantev:2005zs,
    author = "Pantev, Tony G. and Eric Redmund Sharpe",
    title = "{GLSM}s for Gerbes (and other toric stacks)",
    eprint = "hep-th/0502053",
    archivePrefix = "arXiv",
    doi = "10.4310/ATMP.2006.v10.n1.a4",
    journal = "Advances in Theoretical and Mathematical Physics",
    volume = "10",
    number = "1",
    pages = "77--121",
    year = "2006"
}

@misc{Pantev:2005rh,
    author = "Pantev, Tony G. and Eric Redmund Sharpe",
    title = "Notes on gauging noneffective group actions",
    eprint = "hep-th/0502027",
    archivePrefix = "arXiv",
    month = feb,
    year = "2005",
    doi = {10.48550/arXiv.hep-th/0502027},
}

@article{Bucher:1991bc,
    author = "Bucher, Martin Aaron and Lee, Kai-Ming and Preskill, John Phillip",
    title = "On detecting discrete {C}heshire charge",
    eprint = "hep-th/9112040",
    archivePrefix = "arXiv",
    reportNumber = "CALT-68-1753",
    doi = "10.1016/0550-3213(92)90174-A",
    journal = "Nuclear Physics~B",
    volume = "386",
    pages = "27--42",
    year = "1992",
    number = 1,
    month = nov,
}

@article{Alford:1992yx,
    author = "Alford, Mark Gower and Lee, Kai-Ming and March-Russell, John David and Preskill, John Phillip",
    title = "Quantum field theory of non-abelian strings and vortices",
    eprint = "hep-th/9112038",
    archivePrefix = "arXiv",
    reportNumber = "PUPT-91-1288, CALT-68-1700, NSF-ITP-91-128",
    doi = "10.1016/0550-3213(92)90468-Q",
    journal = "Nuclear Physics~B",
    volume = "384",
    pages = "251--317",
    year = "1992",
    month = oct,
    number = {1--2},
}

@article{Alford:1990fc,
    author = "Alford, Mark Gower and March-Russell, John David",
    title = "New order parameters for non-abelian gauge theories",
    reportNumber = "NSF-ITP-90-231, PUPT-1226",
    doi = "10.1016/0550-3213(92)90387-Q",
    journal = "Nuclear Physics~B",
    volume = "369",
    number = {1--2},
    pages = "276--298",
    year = "1992",
    month = jan,
}

@inproceedings{deWildPropitius:1995hk,
    author = "de Wild Propitius, Mark and Bais, Ferdinand Alexander",
    title = "Discrete gauge theories",
    booktitle = "Particles and Fields",
    series = { CRM Series in Mathematical Physics},
    eprint = "hep-th/9511201",
    archivePrefix = "arXiv",
    reportNumber = "PAR-LPTHE-95-46, ITFA-95-20",
    pages = "353--439",
    month = dec,
    year = "1998",
    doi = {10.1007/978-1-4612-1410-6_8},
    publisher = {Springer},
    editor = {Gordon Walter Semenoff and Luc Vinet},
}

@article{Nussinov:2009zz,
    author = "Nussinov, Zohar and Ortíz, Gerardo",
    title = "A symmetry principle for topological quantum order",
    eprint = "cond-mat/0702377",
    archivePrefix = "arXiv",
    doi = "10.1016/j.aop.2008.11.002",
    journal = "Annals of Physics",
    volume = "324",
    number = 5,
    pages = "977--1057",
    year = "2009",
    month = may,
}

@misc{Rist:2022hci,
    author = "Rist, Dominik and Sämann, Christian and Wolf, Martin",
    title = "Explicit Non-{A}belian Gerbes with Connections",
    eprint = "2203.00092",
    archivePrefix = "arXiv",
    primaryClass = "hep-th",
    reportNumber = "EMPG-22-02, DMUS-MP-22/01",
    month = feb,
    year = "2022",
    doi = {10.48550/arXiv.2203.00092},
}

@article{Gaiotto:2014kfa,
    author = "Davide Silvano Achille Gaiotto  and Kapustin, Anton Nikolaevich and Seiberg, Nathan and Willett, Brian",
    title = "Generalized Global Symmetries",
    eprint = "1412.5148",
    archivePrefix = "arXiv",
    primaryClass = "hep-th",
    doi = "10.1007/JHEP02(2015)172",
    journal = "Journal of High Energy Physics",
    volume = 2015,
    number = "02",
    pages = "172",
    year = "2015",
    month = feb,
}

@article{Nguyen:2024ikq,
    author = {Nguy{\~{\^{e}}}n, Mendel T. and Sulejmanpa\v{s}i\'{c}, Tin and {\"U}nsal, Mithat},
    title = "Phases of Theories with \(\mathbb{{Z}}_{{N}}\) 1-Form Symmetry, and the Roles of Center Vortices and Magnetic Monopoles",
    eprint = "2401.04800",
    archivePrefix = "arXiv",
    primaryClass = "hep-th",
    doi = "10.1103/PhysRevLett.134.141902",
    journal = "Physical Review Letters",
    volume = "134",
    number = "14",
    pages = "141902",
    year = "2025",
    month = apr,
}

@article{Sulejmanpasic:2019ytl,
    author = "Sulejmanpa\v{s}i\'{c}, Tin and Gattringer, Christof",
    title = "Abelian gauge theories on the lattice: $\theta$-Terms and compact gauge theory with(out) monopoles",
    eprint = "1901.02637",
    archivePrefix = "arXiv",
    primaryClass = "hep-lat",
    doi = "10.1016/j.nuclphysb.2019.114616",
    journal = "Nuclear Physics~B",
    volume = "943",
    pages = "114616",
    year = "2019",
    month = jun,
}

@article{Gorantla:2021svj,
    author = "Gorantla, Pranay and Lam, Ho Tat and Seiberg, Nathan and Shao, Shu-Heng",
    title = "A modified {V}illain formulation of fractons and other exotic theories",
    eprint = "2103.01257",
    archivePrefix = "arXiv",
    primaryClass = "cond-mat.str-el",
    doi = "10.1063/5.0060808",
    journal = "Journal of Mathematical Physics",
    volume = "62",
    number = "10",
    pages = "102301",
    year = "2021",
    month = oct,
}

@incollection{Gukov:2014gja,
    author = "Gukov, Sergei Gennadievich",
    editor = {Teschner, J\"org},
    title = "Surface operators",
    booktitle = "New Dualities of Supersymmetric Gauge Theories",
    eprint = "1412.7127",
    archivePrefix = "arXiv",
    primaryClass = "hep-th",
    doi = "10.1007/978-3-319-18769-3_8",
    pages = "223--259",
    year = "2015",
    month = nov,
    series = { Mathematical Physics Studies },
    isbn = {978-3-319-18768-6},
    address = {Cham, Switzerland},
    publisher = {Springer},
    issn = { 0921-3767 },
}

@misc{Gukov:2013zka,
    author = "Gukov, Sergei Gennadievich and Kapustin, Anton Nikolaevich",
    title = "Topological Quantum Field Theory, Nonlocal Operators, and Gapped Phases of Gauge Theories",
    eprint = "1307.4793",
    archivePrefix = "arXiv",
    primaryClass = "hep-th",
    month = jul,
    year = "2013",
    doi = {10.48550/arXiv.1307.4793}
}

@incollection{Holland:2000uj,
    author = "Holland, Kieran Michael and Wiese, Uwe-Jens",
    editor = "Shifman, Mikhail Arkadyevich",
    title = "The center symmetry and its spontaneous breakdown at high temperatures",
    booktitle = {At the Frontier of Particle Physics. Boris Ioffe Fest\-schrift},
    eprint = "hep-ph/0011193",
    archivePrefix = "arXiv",
    reportNumber = "BUTP-2000-34, MIT-CTP-3045",
    doi = "10.1142/9789812810458_0040",
    pages = "1909--1944",
    month = apr,
    year = "2001",
    publisher = {World Scientific},
    address = {Singapore},
}

@incollection{Borsten:2024gox,
    author = "Borsten, Leron and Jalali Farahani, Mehran and Jurčo, Branislav and Kim, Hyungrok and Jiří Nárožný and Rist, Dominik and Sämann, Christian and Wolf, Martin",
    title = "Higher Gauge Theory",
    booktitle = "Encyclopedia of Mathematical Physics. Volume 4",
    pages = "159-185",
    edition = "second",
    editor = "Richard Joseph Szabo and Martin Bojowald",
    eprint = "2401.05275",
    archivePrefix = "arXiv",
    primaryClass = "hep-th",
    reportNumber = "EMPG-24-01, DMUS-MP-24/01",
    year = "2025",
    publisher = "Academic Press",
    address = "Cambridge, Massachusetts, United States of America",
    doi = "10.1016/B978-0-323-95703-8.00217-2",
}

@book{hatcher,
 author = {Hatcher, Allen Edward},
 title = {Algebraic Topology},
 isbn = {0-521-79540-0},
 year = {2001},
 month = dec,
 publisher = {Cambridge University Press},
 address = {Cambridge, United Kingdom},
 language = {English},
 zbMATH = {2103273},
 Zbl = {1044.55001},
 url = {https://pi.math.cornell.edu/~hatcher/AT/ATpage.html},
}

@article{Ogilvie:2012is,
    author = "Ogilvie, Michael C.",
    title = "Phases of Gauge Theories",
    eprint = "1211.2843",
    archivePrefix = "arXiv",
    primaryClass = "hep-th",
    doi = "10.1088/1751-8113/45/48/483001",
    journal = "Journal of Physics~A",
    volume = "45",
    number = 48,
    pages = "483001",
    year = "2012",
    month = dec,
}

@misc{Giansiracusa:2025hfj,
    author = "Giansiracusa, Jeffrey H. and Lanners, David and Sulejmanpa\v{s}i\'{c}, Tin",
    title = "Emergent photons and mechanisms of confinement",
    eprint = "2505.00079",
    archivePrefix = "arXiv",
    primaryClass = "hep-lat",
    month = apr,
    year = "2025",
    doi={10.48550/arXiv.2505.00079},
}

@article{Barkeshli:2014cna,
    author = "Barkeshli, Maissam and Bonderson, Parsa Hassan and Cheng, Meng and Wang, Zhenghan",
    title = "Symmetry Fractionalization, Defects, and Gauging of Topological Phases",
    eprint = "1410.4540",
    archivePrefix = "arXiv",
    primaryClass = "cond-mat.str-el",
    doi = "10.1103/PhysRevB.100.115147",
    journal = "Physical Review~B",
    volume = "100",
    number = "11",
    pages = "115147",
    year = "2019",
    month = sep,
}

@incollection{Kapustin:2013uxa,
    author = "Kapustin, Anton Nikolaevich and Thorngren, Ryan",
    title = "Higher Symmetry and Gapped Phases of Gauge Theories",
    booktitle = { Algebra, Geometry, and Physics in the 21st Century: Kontsevich Festschrift},
    editor = { Denis Auroux and Ludmil Vasilev Katzarkov and Tony G. Pantev and {\relax Ya}n Sergeevich Soibelman and {\relax Yu}ri Tschinkel},
    eprint = "1309.4721",
    archivePrefix = "arXiv",
    primaryClass = "hep-th",
    doi = "10.1007/978-3-319-59939-7_5",
    series = "Progress in Mathematics",
    volume = "324",
    pages = "177--202",
    year = "2017",
    isbn= {978-3-319-59938-0},
    month = aug,
    publisher = {Birkhäuser},
    address = {Cham, Switzerland},
}

@article{Breen:2001ie,
    author = "Breen, Lawrence and Messing, William",
    title = "Differential geometry of gerbes",
    eprint = "math/0106083",
    archivePrefix = "arXiv",
    doi = "10.1016/j.aim.2005.06.014",
    journal = "Advances in Mathematics",
    volume = "198",
    number = 2,
    pages = "732-846",
    year = "2005",
    month = dec,
}

@article{Aschieri:2003mw,
    author = "Aschieri, Paolo and Cantini, Luigi and Jurčo, Branislav",
    title = "Nonabelian bundle gerbes, their differential geometry and gauge theory",
    eprint = "hep-th/0312154",
    archivePrefix = "arXiv",
    reportNumber = "MPP-2003-139, LMU-TPW-07-03",
    doi = "10.1007/s00220-004-1220-6",
    journal = "Communications in Mathematical Physics",
    volume = "254",
    pages = "367--400",
    year = "2005",
    month = mar,
}

@misc{Baez:2004in,
    author = "Baez, John Carlos and Schreiber, Urs",
    title = "Higher gauge theory: 2-connections on 2-bundles",
    eprint = "hep-th/0412325",
    archivePrefix = "arXiv",
    month = dec,
    year = "2004",
    doi = {10.48550/arXiv.hep-th/0412325},
}

@incollection{Baez:2005qu,
    author = "Baez, John Carlos and Schreiber, Urs",
    title = "Higher gauge theory",
    eprint = "math/0511710",
    archivePrefix = "arXiv",
    year = "2007",
    booktitle = "Categories in Algebra, Geometry and Mathematical Physics. Conference and Workshop in Honor of Ross Street's 60th Birthday, July 11--16/July 18--21, 2005, Macquarie University, Sydney, Australia; Australian National University, Canberra, Australia",
    publisher = "American Mathematical Society",
    address = "Providence, Rhode Island, United States of America",
    series = "Contemporary Mathematics",
    volume = 431,
    pages = "7--31",
    isbn = "978-0-8218-3970-6",
    editor = "Alexei Alexandrovich Davydov and Michael Alexandrovich Batanin and Michael Sterling James Johnson and Stephen Lack and Amnon Neeman",
    doi = "10.1090/conm/431/08264",
}

@article{Samann:2019eei,
    author = {S\"amann, Christian and Schmidt, Lennart},
    title = "Towards an {M}5-Brane Model {II}: Metric String Structures",
    eprint = "1908.08086",
    archivePrefix = "arXiv",
    primaryClass = "hep-th",
    reportNumber = "EMPG-19-20",
    doi = "10.1002/prop.202000051",
    journal = "\foreignlanguage{german}{Fortschritte der Physik}",
    volume = "68",
    number = "8",
    pages = "2000051",
    year = "2020",
    month = aug,
}

@article{Gastel:2018joi,
    author = "Gastel, Andreas",
    title = "Canonical gauges in higher gauge theory",
    eprint = "1810.06278",
    archivePrefix = "arXiv",
    primaryClass = "math-ph",
    doi = "10.1007/s00220-019-03530-4",
    journal = "Communications in Mathematical Physics",
    volume = "376",
    number = "2",
    pages = "1053--1071",
    year = "2020",
    month = jun,
}

@article{Schmidt:2019pks,
    author = "Schmidt, Lennart",
    title = "Twisted {W}eil Algebras for the String {L}ie 2-Algebra",
    eprint = "1903.02873",
    archivePrefix = "arXiv",
    primaryClass = "hep-th",
    doi = "10.1002/prop.201910016",
    journal = "\foreignlanguage{german}{Fortschritte der Physik}",
    volume = "67",
    number = "8--9",
    pages = "1910016",
    year = "2019",
    month = aug # "--" # sep,
}

@article{Kim:2019owc,
    author = "Kim, Hyungrok and Sämann, Christian",
    title = "Adjusted parallel transport for higher gauge theories",
    eprint = "1911.06390",
    archivePrefix = "arXiv",
    primaryClass = "hep-th",
    reportNumber = "EMPG-19-24",
    doi = "10.1088/1751-8121/ab8ef2",
    journal = "Journal of Physics~A",
    volume = "53",
    number = "44",
    pages = "445206",
    year = "2020",
    month = nov,
}

@misc{Borsten:2021ljb,
    author = "Borsten, Leron and Kim, Hyungrok and Sämann, Christian",
    title = "${EL}_\infty$-algebras, Generalized Geometry, and Tensor Hierarchies",
    eprint = "2106.00108",
    archivePrefix = "arXiv",
    primaryClass = "hep-th",
    reportNumber = "EMPG-21-07",
    month = may,
    year = "2021",
    doi = {10.48550/arXiv.2106.00108},
}

@misc{Fischer:2024vak,
    author = "Fischer, Simon-Raphael and Jalali Farahani, Mehran and Kim, Hyungrok and Sämann, Christian",
    title = "Adjusted Connections {I}: Differential Cocycles for Principal Groupoid Bundles with Connection",
    eprint = "2406.16755",
    archivePrefix = "arXiv",
    primaryClass = "math.DG",
    month = jun,
    year = "2024",
    doi = {10.48550/arXiv.2406.16755},
}

@incollection{Sati:2008eg,
    author = "Sati, Hisham and Schreiber, Urs and James Dillon Stasheff",
    title = "${L}_\infty$-algebra connections and applications to \(\operatorname{String}\)- and {C}hern-{S}imons \(n\)-transport",
    booktitle = { Quantum Field Theory: Competitive Models},
    eprint = "0801.3480",
    archivePrefix = "arXiv",
    primaryClass = "math.DG",
    doi = "10.1007/978-3-7643-8736-5_17",
    pages = "303-424",
    month = dec,
    year = "2009",
    isbn = {978-3-7643-8735-8},
    publisher = {Birkhäuser},
    address = {Basel, Switzerland},
    editor = {Bertfried Fauser and Jürgen Tolksdorf and Eberhard Zeidler},
}

@article{Sati:2009ic,
    author = "Sati, Hisham and Schreiber, Urs and James Dillon Stasheff",
    title = "Differential twisted \(\operatorname{String}\) and \(\operatorname{Fivebrane}\) structures",
    eprint = "0910.4001",
    archivePrefix = "arXiv",
    primaryClass = "math.AT",
    doi = "10.1007/s00220-012-1510-3",
    journal = "Communications in Mathematical Physics",
    volume = "315",
    number = 1,
    pages = "169--213",
    year = "2012",
    month = oct,
}

@article{Bhardwaj:2023kri,
    author = "Bhardwaj, Lakshya and Bottini, Lea E. and Fraser-Taliente, Ludovic and Gladden, Liam and Gould, Dewi Sid William and Platschorre, Arthur and Tillim, Hannah",
    title = "Lectures on generalized symmetries",
    eprint = "2307.07547",
    archivePrefix = "arXiv",
    primaryClass = "hep-th",
    doi = "10.1016/j.physrep.2023.11.002",
    journal = "Physics Reports",
    volume = "1051",
    pages = "1--87",
    year = "2024",
    month = feb,
}

@article{Gomes:2023ahz,
    author = "Pedro Rogério Sérgi Gomes",
    title = "An introduction to higher-form symmetries",
    eprint = "2303.01817",
    archivePrefix = "arXiv",
    primaryClass = "hep-th",
    doi = "10.21468/SciPostPhysLectNotes.74",
    journal = "SciPost Physics Lecture Notes",
    volume = "74",
    pages = "1--56",
    year = "2023",
    month = sep,
}

@misc{Cordova:2022ruw,
    author = "Córdova, Clay and Dumitrescu, Thomas T. and Intriligator, Kenneth A. and Shao, Shu-Heng",
    title = "Snowmass White Paper: Generalized Symmetries in Quantum Field Theory and Beyond",
    eprint = "2205.09545",
    archivePrefix = "arXiv",
    primaryClass = "hep-th",
    month = may,
    year = "2022",
    doi = {10.48550/arXiv.2205.09545},
}

@misc{Brennan:2023mmt,
    author = "Brennan, Theodore Daniel and Hong, Sungwoo",
    title = "Introduction to Generalized Global Symmetries in {QFT} and Particle Physics",
    eprint = "2306.00912",
    archivePrefix = "arXiv",
    primaryClass = "hep-ph",
    month = jun,
    year = "2023",
    doi = {10.48550/arXiv.2306.00912},
}

@article{Luo:2023ive,
    author = "Luo, Ran and Wang, Qing-Rui and Wang, Yi-Nan",
    title = "Lecture notes on generalized symmetries and applications",
    eprint = "2307.09215",
    archivePrefix = "arXiv",
    primaryClass = "hep-th",
    doi = "10.1016/j.physrep.2024.02.002",
    journal = "Physics Reports",
    volume = "1065",
    pages = "1--43",
    year = "2024",
    month = may,
}

@article{Saemann:2017rjm,
    author = "Sämann, Christian and Schmidt, Lennart",
    title = "The non-abelian self-dual string",
    eprint = "1705.02353",
    archivePrefix = "arXiv",
    primaryClass = "hep-th",
    reportNumber = "EMPG-17-05",
    doi = "10.1007/s11005-019-01250-3",
    journal = "Letters in Mathematical Physics",
    volume = "110",
    number = "5",
    pages = "1001--1042",
    year = "2020",
    month = may,
}

@article{Baez:2010ya,
    author = "Baez, John Carlos and Huerta, John",
    title = "An Invitation to Higher Gauge Theory",
    eprint = "1003.4485",
    archivePrefix = "arXiv",
    primaryClass = "hep-th",
    doi = "10.1007/s10714-010-1070-9",
    journal = "General Relativity and Gravitation",
    volume = "43",
    pages = "2335--2392",
    year = "2011",
    month = aug,
}

@article{Villain:1974ir,
    author = "Villain, Jacques",
    title = "Theory of one- and two-dimensional magnets with an easy magnetization plane. {II}. {T}he planar, classical, two-dimensional magnet",
    doi = "10.1051/jphys:01975003606058100",
    journal = "\foreignlanguage{french}{Le Journal de Physique}",
    volume = "36",
    pages = "581--590",
    number = 6,
    year = "1975",
    month = jun,
}

@article{tHooft:1977nqb,
    author = "'t Hooft, Gerardus",
    title = "On the Phase Transition Towards Permanent Quark Confinement",
    reportNumber = "Print-78-0099 (UTRECHT)",
    doi = "10.1016/0550-3213(78)90153-0",
    journal = "Nuclear Physics~B",
    volume = "138",
    number = 1,
    pages = "1--25",
    year = "1978",
    month = jun,
}

@article{Cornwall:1979hz,
    author = "Cornwall, John Michael",
    title = "Quark Confinement and Vortices in Massive Gauge-Invariant {QCD}",
    reportNumber = "UCLA/79/TEP/5",
    doi = "10.1016/0550-3213(79)90111-1",
    journal = "Nuclear Physics~B",
    volume = "157",
    number = 3,
    pages = "392--412",
    year = "1979",
    month = oct,
}

@article{Nielsen:1979xu,
    author = "Nielsen, Holger Bech and Olesen, Poul",
    title = "A Quantum Liquid Model for the {QCD} Vacuum: Gauge and Rotational Invariance of Domained and Quantized Homogeneous Color Fields",
    reportNumber = "NBI-HE-79-17",
    doi = "10.1016/0550-3213(79)90065-8",
    journal = "Nuclear Physics~B",
    volume = "160",
    number = 2,
    pages = "380--396",
    year = "1979",
    month = dec,
}

@article{Hayashi:2024yjc,
    author = "Hayashi, Yui and Tanizaki, Yuya",
    title = "Unifying Monopole and Center Vortex as the Semiclassical Confinement Mechanism",
    eprint = "2405.12402",
    archivePrefix = "arXiv",
    primaryClass = "hep-th",
    reportNumber = "YITP-24-59",
    doi = "10.1103/PhysRevLett.133.171902",
    journal = "Physical Review Letters",
    volume = "133",
    number = "17",
    pages = "171902",
    year = "2024",
    month = oct,
}

@article{Hayashi:2024psa,
    author = "Hayashi, Yui and Misumi, Tatsuhiro and Tanizaki, Yuya",
    title = "Monopole-vortex continuity of \(\mathcal{N}=1\) super {Y}ang-{M}ills theory on \(\mathbb{R}^2\times{S}^1\times{S}^1\) with 't~{H}ooft twist",
    eprint = "2410.21392",
    archivePrefix = "arXiv",
    primaryClass = "hep-th",
    reportNumber = "YITP-24-136",
    doi = "10.1007/JHEP05(2025)194",
    journal = "Journal of High Energy Physics",
    volume = 2025,
    number = "05",
    pages = "194",
    year = "2025",
    month = may,
}

@article{Polyakov:1975rs,
    author = "Polyakov, Alexander Markovich",
    editor = "Taylor, J. C.",
    title = "Compact Gauge Fields and the Infrared Catastrophe",
    doi = "10.1016/0370-2693(75)90162-8",
    journal = "Physics Letters~B",
    volume = "59",
    number = 1,
    pages = "82--84",
    year = "1975",
    month = oct,
}

@article{Polyakov:1976fu,
    author = "Polyakov, Alexander Markovich",
    title = "Quark Confinement and Topology of Gauge Groups",
    reportNumber = "NORDITA-76/33",
    doi = "10.1016/0550-3213(77)90086-4",
    journal = "Nuclear Physics~B",
    volume = "120",
    number = 3,
    pages = "429--458",
    year = "1977",
    month = mar,
}

@article{getzlerjonespetrack,
 author = {Getzler, Ezra and Jones, John David Stuart and Petrack, Scott B.},
 title = {Differential forms on loop spaces and the cyclic bar complex},
 journal = {Topology},
 issn = {0040-9383},
 volume = {30},
 number = {3},
 pages = {339--371},
 year = {1991},
 language = {English},
 doi = {10.1016/0040-9383(91)90019-Z},
 zbMATH = {4205422},
 Zbl = {0729.58004},
 month = nov,
}

@misc{Tellez-Dominguez:2023wwr,
    author = "Téllez-Domínguez, Roberto",
    title = "Chern correspondence for higher principal bundles",
    eprint = "2310.12738",
    archivePrefix = "arXiv",
    primaryClass = "math.DG",
    month = oct,
    year = "2023",
    doi = {10.48550/arXiv.2310.12738},
}

@misc{Gagliardo:2025oio,
    author = "Gagliardo, Gianni and Sämann, Christian and Téllez-Domínguez, Roberto",
    title = "Principal 3-Bundles with Adjusted Connections",
    eprint = "2505.13368",
    archivePrefix = "arXiv",
    primaryClass = "math-ph",
    month = may,
    year = "2025",
    doi = {10.48550/arXiv.2505.13368}
}

@misc{future:2024aa,
   author = "Simon-Raphael Fischer and Jalali Farahani, Mehran and Hyungrok Kim and Christian Sämann",
   title  = "Adjusted connections II: Differential cocycles for higher principal bundles with connections",
   note   = "To appear"
}

@article{zbMATH03361026,
 author = {Chen, Kuo-Tsai},
 title = {Iterated integrals of differential forms and loop space homology},
 journal = {Annals of Mathematics},
 issn = {0003-486X},
 volume = {97},
 number = 2,
 month = mar,
 pages = {217--246},
 year = {1973},
 language = {English},
 doi = {10.2307/1970846},
 zbMATH = {3361026},
 Zbl = {0227.58003}
}

@inproceedings{zbMATH03532253,
 author = {Chen, Kuo-Tsai},
 title = {Reduced bar constructions on de{R}ham complexes},
 year = {1976},
 language = {English},
 booktitle = {Algebra, Topology, and Category Theory: A Collection of Papers in Honor of Samuel Eilenberg},
 editor = {Alex Heller and Myles Tierney},
 doi = {10.1016/B978-0-12-339050-9.50007-9},
 zbMATH = {3532253},
 Zbl = {0341.57034},
 publisher = {Academic Press},
 address = {New York, United States of America},
}

@article{zbMATH04050538,
 author = {Jones, John David Stuart},
 title = {Cyclic homology and equivariant homology},
 journal = {\foreignlanguage{latin}{Inventiones Mathematicae}},
 issn = {0020-9910},
 volume = {87},
 number = 2,
 pages = {403--423},
 year = {1987},
 month = jun,
 language = {English},
 doi = {10.1007/BF01389424},
 zbMATH = {4050538},
 Zbl = {0644.55005}
}

@article{Banks:2010zn,
    author = "Banks, Thomas Israel and Seiberg, Nathan",
    title = "Symmetries and Strings in Field Theory and Gravity",
    eprint = "1011.5120",
    archivePrefix = "arXiv",
    primaryClass = "hep-th",
    doi = "10.1103/PhysRevD.83.084019",
    journal = "Physical Review~D",
    volume = "83",
    number = 8,
    pages = "084019",
    year = "2011",
    month = apr,
}

@article{Orland:1981ku,
    author = "Orland, Peter",
    title = "Instantons and disorder in antisymmetric tensor gauge fields",
    reportNumber = "UCSC/81/141",
    doi = "10.1016/0550-3213(82)90468-0",
    journal = "Nuclear Physics B",
    volume = "205",
    number = 1,
    pages = "107--118",
    year = "1982",
    month = apr,
}

\end{document}